\documentclass[floatfix,twocolumn,showkeys,preprintnumbers,amsmath,amssymb]{revtex4}

\usepackage{graphicx}
\usepackage{dcolumn}
\usepackage{bm}

\begin{document}


\title{Atomic ordering effects and bond indices in quaternary systems
Si-(Be,Al)-O-N}

\author{S.V. Okatov}
\email[]{okatov@ihim.uran.ru}
\affiliation{Institute of Solid State Chemistry, 
Ural Branch of the Russian Academy of Sciences, Pervomayskaya st.,
GSP-145, 620219, Ekaterinburg, Russia}

\date{\today}

\begin{abstract}
We propose a microscopic method for modeling atomic ordering effects in
multi-component solid solutions, which is based on minimizing the number
of unfavorable interatomic bonds in the system. Atomic ordering effects
are studied in the solid solutions Si$_{6-x}$Al$_x$O$_x$N$_{8-x}$
($\beta$-SiAlONs) and hypothetical Si$_{6-x}$Be$_x$O$_{2x}$N$_{8-2x}$ in
the $\beta$-Si$_3$N$_4$-Be-O system. It is established that in
$\beta$-SiAlONs, Al,O atoms form separate quasi-one-dimensional
``channels''. On the contrary, in Si$_{6-x}$Be$_x$O$_{2x}$N$_{8-2x}$, Be
and O atoms constitute a whole cluster. This means that homogeneous
solid solutions cannot be formed in the $\beta$-Si$_3$N$_4$-Be-O system
under equilibrium conditions.

\end{abstract}



\maketitle

\section{Introduction}

Atomic ordering effects (AOE) have been discovered in a wide range of
non-stoi\-chio\-met\-ric and multi-component alloys, solid solutions
(SS) and compounds. The influence they produce on the properties is
great and comparable with the effect of altering the chemical
composition of materials \cite{gus}.

In the majority of theoretical simulations of multi-component materials,
atomic ordering effects are studied using integral energy
characteristics (for example, free, total or cohesion energies) of a
limited number of atomic configurations of a system set {\it a priory}
\cite{order}. Generally, this approach is applied to binary or ternary
compounds, whereas for more complex systems its application is confined
by computer power due to rapidly increasing sizes of cells to be
considered and the number of atomic configurations.

In the present report, we describe a new method for studying AOE based
on comparative analysis of detached types of interatomic bonds in
multi-component systems. In the framework of this method, optimal atomic
configurations are chosen by minimizing unfavorable interatomic bonds.

The method is illustrated using quaternary solid solutions
Si$_{6-x}$Al$_x$O$_x$N$_{8-x}$ ($\beta$-SiAlONs) considered earlier in
ref. \cite{okatov2001} as an example. It is also applied to predict
atomic ordering effects in the hypothetical $\beta$-Si$_3$N$_4$-based
solid solutions Si$_{6-x}$Be$_x$O$_{2x}$N$_{8-2x}$.

\section{The method for modeling atomic ordering effects:
$\beta$-SiAlONs}

As is known, the stability of a thermodynamic system corresponds to the
free energy or, at zero temperature, total energy minimum condition. In
terms of the chemical bonding theory, this condition implies the optimal
type of bands filling, when all bonding states are occupied and all
antibonding states are vacant.

In the proposed method, AOE are determined by searching for the
multi-component system configurations having the maximum number of the
most favorable interatomic bonds. To characterize the detached type of
paired bonds, crystal orbital overlap populations (COOP) and the numbers
of filled antibonding states (FAS) may be used. In this work, these
values are calculated employing the Mulliken analysis by the
tight-binding band method with matrix elements parameterization
according to the H\"uckel theory \cite{huck}. Other techniques of bond
characterization can be used as well.

The application of this approach can be demonstrated on the
four-component SS in the Si-Al-O-N system with the basic structure
$\beta$-Si$_3$N$_4$ (the so-called $\beta$-SiAlONs of the formal
stoichiometry Si$_{6-x}$Al$_x$O$_x$N$_{8-x}$, where $x=0-4.2$).

\begin{figure}[t]
\resizebox*{\columnwidth}{!}{\rotatebox{0}{\includegraphics{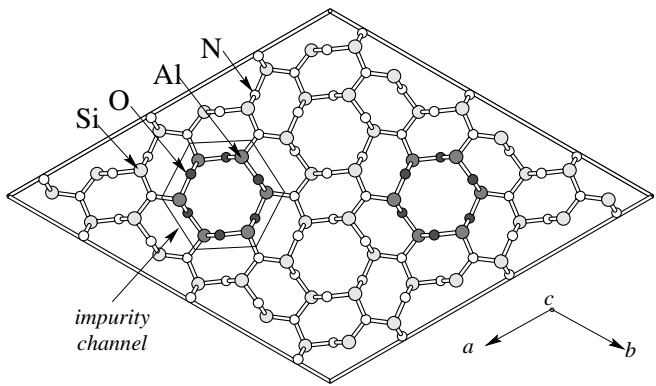}}}
\caption{
The impurity channels in the structure of Si$_{6-x}$Al$_x$O$_x$N$_{8-x}$
($x=1.3$).}
\label{f.sial_struct}
\end{figure}

A previous investigation of $\beta$-SiAlONs by the tight-binding band
method \cite{okatov2001} showed that Al and O atoms form extended
quasi-one-dimensional (1D) structures constituted by 12-atomic rings ---
the so-called ``impurity channels'', fig. \ref{f.sial_struct}.
Subsequent studies by the DFT cluster \cite{ryz} and the band structure
full-potential LMTO methods \cite{flmto} showed that ordered structures
exhibited the best stability. Besides, their chemical bonding and
electronic properties were examined.

A detailed analysis of bond indices in $\beta$-SiAlONs demonstrated
\cite{okatov2001} that the most unfavorable bonds are Si-O, which
contain the maximum number of FAS. The antibonding states of Al-O bonds
are less populated. Si-N and Al-N bonds do not contain any FAS, while
COOP(Si-N$=0.58$ e)$>$COOP(Al-N$=0.24$ e). Hence, a bonds hierarchy
taking into account their stabilizing effects can be represented by the
so-called sequence of bonds preferences:

\begin{equation}
P_{\rm Si-O} < P_{\rm Al-O} < P_{\rm Al-N} < P_{\rm Si-N},
\label{eq1}
\end{equation}
where $P_{\rm A-B}$ reflects the preference of A-B-type bonds presence
in the system.

Thus, in order to determine the superstructure of a multi-component
material, it is necessary to choose such distributions of atoms over
lattice sites, in which the number of bonds is minimized consecutively
in accordance with (\ref{eq1}).

For example, for $\beta$-SiAlONs at the first step the superstructures
with the minimum number of Si-O bonds are chosen from all possible ones
according to (\ref{eq1}). Among them we choose the superstructures with
the minimum number of Al-O bonds etc. As a result, in a system of a
given composition, the superstructures are found, which possess the same
number of the same types of bonds, but differ only by their positions in
the supercell.

To determine the atomic order type, the following algorithm is applied.
The SS Si$_{6-x}$Al$_x$O$_x$N$_{8-x}$ of a certain composition modelled
by a supercell is considered. In this work we make use of the 126-atomic
supercells (3x3x1) analogous to those applied in \cite{okatov2001}. For
example, the composition of the SS with $x=2.0$ (Si$_4$Al$_2$O$_2$N$_6$)
is described by the Si$_{36}$Al$_{18}$O$_{18}$N$_{54}$ cell. The
following items are fulfilled:

\begin{description}
\item[(i)] Atoms (Si,Al,N,O) are randomly distributed over the supercell
sites so that the result does not depend on initial conditions.

\item[(ii)] A random substitution, which is the only permitted one
for this system (for example Si$\rightarrow$Al, N$\rightarrow$O), is
performed, while all other types of substitution (Si$\rightarrow$O,
N$\rightarrow$Al) are not considered.

\item[(iii)] If the correspondence to (\ref{eq1}) improves (the number
of favorable bonds increases and the number of unfavorable bonds3
decreases), the step is accepted, otherwise it is rejected.

\item[(iv)] Items (ii) and (iii) are repeated until the number of steps
denied in succession does not exceed $N_1$.

\item[(v)] Items (i-iv) are repeated until the number of structures
denied in succession does not exceed $N_2$. This cycle is necessary
since the result of cycle (iv) depends on the initial position of atoms
and on the first accepted substitution (item iii) defining the
subsequent shape of the superstructure.
\end{description}

The choice of $N_1$, $N_2$ values is determined by the supercell size
and composition. In the present work they range from 10000 to 100000 and
from 10 to 400 respectively.

An application of the above algorithm for $\beta$-SiAlONs with the same
cells and compositions as in \cite{okatov2001} shows that the resulting
superstructures shapes (1D impurity channels, fig. \ref{f.sial_struct})
are in complete agreement with those obtained in a series of
complicated calculations of total energies. The formation of impurity
channels in $\beta$-SiAlONs leads to the disappearance of the most
unfavorable Si-O bonds, while the number of Al-O bonds is minimal.

The evident drawback of the aforesaid scheme concerns condition
(\ref{eq1}), which reflects only the effect of paired bonds A-B. It
should not be excluded that in some cases the formation of A$_n$B$_m$
clusters is more preferable in spite of low preference of A-B bonds.
This restriction may be removed by using a sequence of clusters
preferences instead of a sequence of bonds preferences. In respect to
$\beta$-SiAlONs, the above restriction does not permit differentiating
between distant and adjacent positions of impurity channels.

\section{Electronic structure and atomic ordering effects in the
$\beta$-Si$_3$N$_4$-Be-O system}

\subsection{Bond indices and chemical composition}

\begin{figure*}[t]
\resizebox*{\textwidth}{!}{\rotatebox{270}{\includegraphics{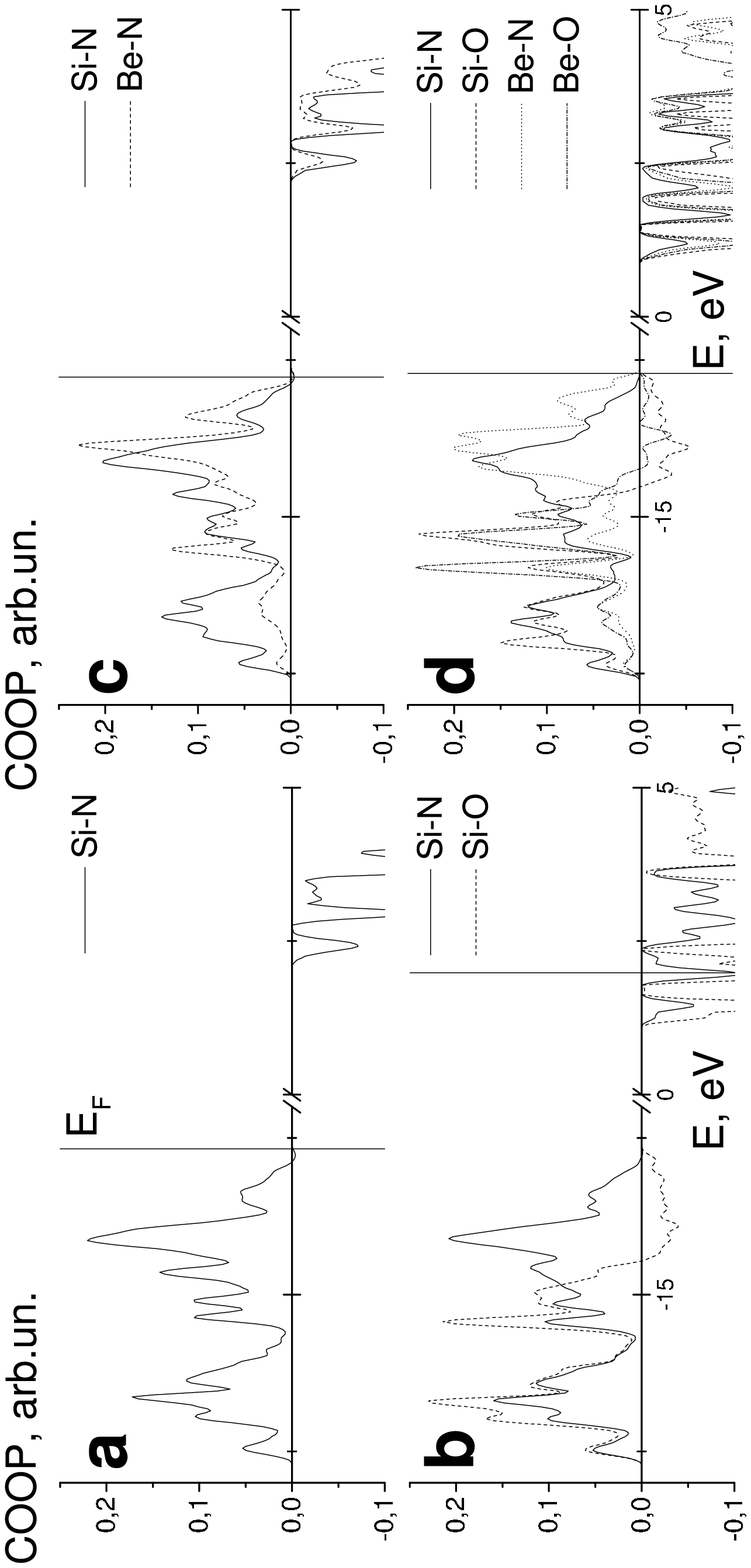}}}
\caption{Crystal orbital overlap populations of bonds in Si$_3$N$_4$ (a)
and impurity systems Si$_3$N$_4$:O (b), Si$_3$N$_4$:Be (c),
Si$_3$N$_4$:(Be+2O) (d) simulated by Si$_{54}$N$_{72}$,
Si$_{54}$ON$_{71}$, Si$_{53}$BeN$_{72}$ É Si$_{53}$BeO$_{2}$N$_{70}$
cells, respectively.}
\label{f.bonds}
\end{figure*}

At the first stage of modeling the hypothetical SS in the quaternary
system $\beta$-Si$_3$N$_4$-Be-O, we analyzed the electronic properties
and bond indices of pure $\beta$-Si$_3$N$_4$ and that containing Be and
O impurities, fig. \ref{f.bonds}. It is seen that in the binary nitride
$\beta$-Si$_3$N$_4$ (cell Si$_{54}$N$_{72}$, valence electrons
concentration (VEC) equals 576 e/cell), all antibonding states are
vacant, while all bonding states are completely occupied.

O$\rightarrow$N or Be$\rightarrow$Si substitutions (cells
Si$_{54}$ON$_{71}$ and Si$_{53}$BeN$_{72}$ respectively) reduce the
stability of the $\beta$-Si$_3$N$_4$:O and $\beta$-Si$_3$N$_4$:Be
systems. In the first case this is due to partial filling of antibonding
states when VEC increases (577 e/cell), while in the second case --- due
to a depopulation some bonding states when VEC decreases (574 e/cell). A
``mutual compensation'' of the destabilizing effects of single
impurities Be, O can be achieved by choosing a composition, at which the
VEC remains 576 e/cell, i.e. when the impurities are introduced in the
proportion (Be+2O). In this case the type of bands filling (as compared
with the basic $\beta$-Si$_3$N$_4$) does not change and the number of
filled antibonding and empty bonding states is minimal, fig.
\ref{f.bonds}. Thus, the formal stoichiometry of the SS in the
$\beta$-Si$_3$N$_4$-Be-O system is Si$_{6-x}$Be$_x$O$_{2x}$N$_{8-2x}$.

\subsection{Short-range atomic ordering}

When heterovalent substitutions occur in $\beta$-Si$_3$N$_4$, the
adjacent arrangement of donour (O) and acceptor (Be) atoms (i.e. the
formation of \{BeO$_2$\} complexes) is more probable, otherwise the
charge screening effect hinders the above mentioned compensation of
bonds filling.

\subsection{Long-range atomic ordering}

Analysis of separate bonds in the Si$_{6-x}$Be$_x$O$_{2x}$N$_{8-2x}$
system shows that Si-O and Be-O bonds consist of filled antibonding
states, and their number in Si-O is greater than in Be-O (fig.
\ref{f.bonds}). Determining atomic ordering effects by the proposed
algorithm, we search for atomic configurations satisfying the condition
of type (\ref{eq1}):

\begin{equation}
P_{\rm Si-O} < P_{\rm Be-O} < P_{\rm Be-N} < P_{\rm Si-N},
\end{equation}
In other words, for the considered series of solid solutions of variable
concentrations Si$_{6-x}$Be$_x$O$_{2x}$N$_{8-2x}$ ($x\approx 0-0.5$,
supercells Si$_{53}$BeO$_2$N$_{70}$ --
Si$_{24}$Be$_{30}$O$_{60}$N$_{12}$), we determine the configurations of
modelled cells, in which the number of less advantageous bonds Si-O and
Be-O is successively minimized.

\begin{figure*}[t]
\resizebox*{\textwidth}{!}{\rotatebox{0}{\includegraphics{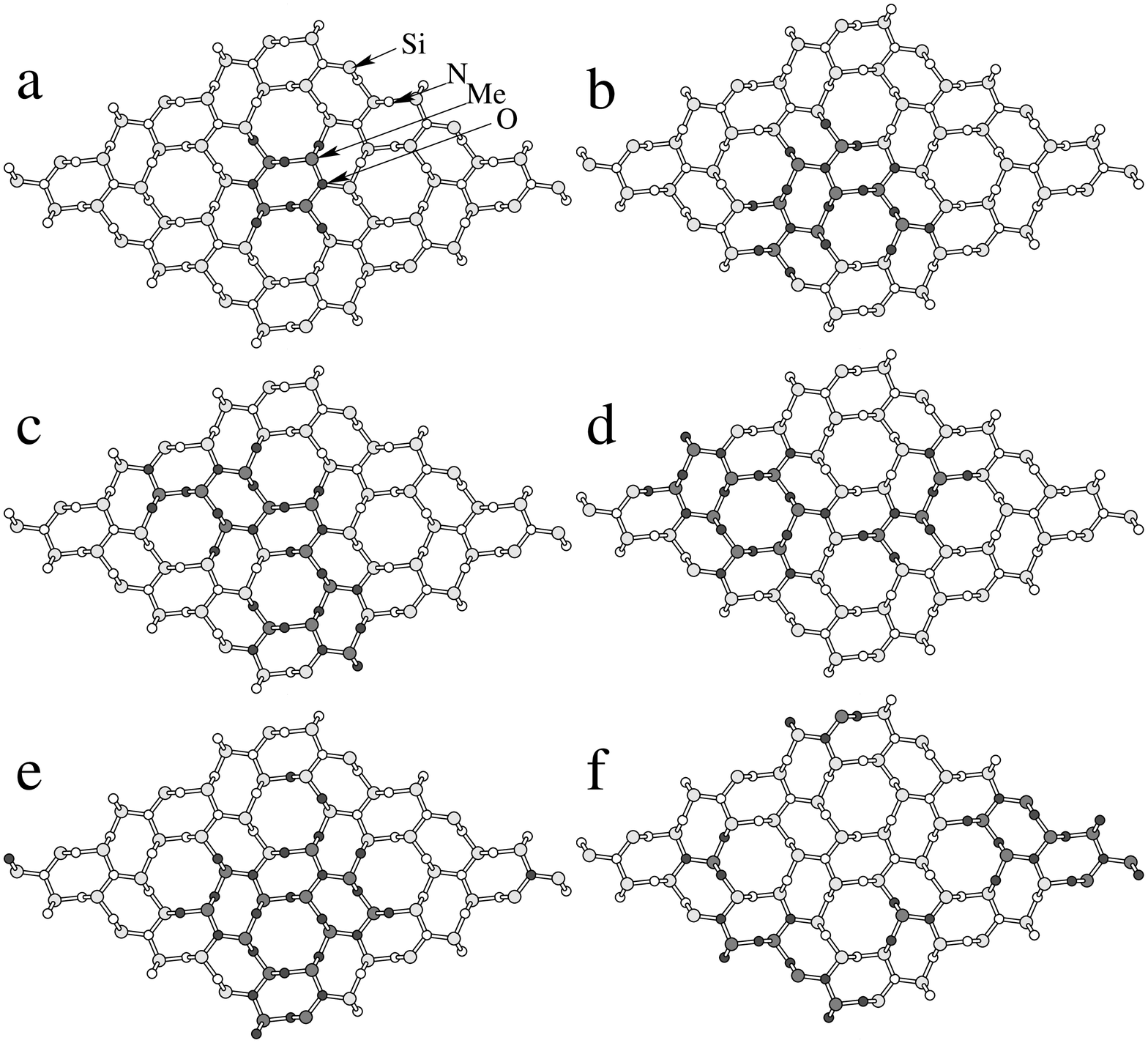}}}
\caption{Superstructures in cells of
Si$_{50}$Be$_4$O$_8$N$_{64}$ (a),  Si$_{46}$Be$_8$O$_{16}$N$_{58}$ (b),
Si$_{43}$Be$_{11}$O$_{22}$N$_{50}$ (c,d),
Si$_{41}$Be$_{13}$O$_{26}$N$_{46}$ (e),
Si$_{40}$Be$_{14}$O$_{28}$N$_{44}$ (f) compositions.}
\label{f.sibeon_struct}
\end{figure*}

The results of the simulation are depicted in fig. \ref{f.sibeon_struct}
and table \ref{t.n_bonds}. Their major difference from
$\beta$-Si$_{6-x}$Al$_x$O$_x$N$_{8-x}$ is the presence of Si-O bonds,
the number of which grows monotonously with the concentration of Be and
O. This indicates that Si$_{6-x}$Be$_x$O$_{2x}$N$_{8-2x}$ is less stable
in comparison with Si$_{6-x}$Al$_x$O$_x$N$_{8-x}$. When $x$ is small
(cells Si$_{53}$BeO$_2$N$_{70}$ -- Si$_{51}$Be$_3$O$_6$N$_{66}$), the
system contains Be-N bonds, and the number of ``unfavorable'' Si-O bonds
grows with $x$ most rapidly. The noted peculiarities are due to a small
number of Be and O atoms in the supercell and its limited size. In real
crystals this situation is not possible.

\begin{table}[b]
\caption{The numbers of different-type bonds in hypothetical SS
Si$_{6-x}$Be$_x$O$_{2x}$N$_{8-2x}$ as a function of the modelled cell
composition.}
\centering
\begin{tabular*}{\columnwidth}{p{2.5cm}|c@{\extracolsep{\fill}}cccc}
\hline
&\multicolumn{5}{c}{Bonds}\\
\cline{2-6}
Cell&&Si-O&Be-O&Be-N&Si-N\\
\hline
Si$_{53}$BeO$_2$N$_{70}$	&&3	&3	&1	&209\\
Si$_{52}$Be$_2$O$_4$N$_{68}$	&&5	&7	&1	&203\\
Si$_{51}$Be$_3$O$_6$N$_{66}$	&&7	&11	&1	&197\\
Si$_{50}$Be$_4$O$_8$N$_{64}$	&&8	&16	&0	&192\\
Si$_{49}$Be$_5$O$_{10}$N$_{62}$	&&10	&20	&0	&186\\
Si$_{48}$Be$_6$O$_{12}$N$_{60}$	&&12	&24	&0	&180\\
Si$_{47}$Be$_7$O$_{14}$N$_{58}$	&&14	&28	&0	&174\\
Si$_{46}$Be$_8$O$_{16}$N$_{56}$	&&16	&32	&0	&168\\
\multicolumn{5}{c}{....................}\\
Si$_{26}$Be$_{28}$O$_{56}$N$_{16}$	&&56	&112	&0	&48\\
Si$_{25}$Be$_{29}$O$_{58}$N$_{14}$	&&58	&116	&0	&42\\
Si$_{24}$Be$_{30}$O$_{60}$N$_{12}$	&&60	&120	&0	&36\\
\hline
\end{tabular*}
\label{t.n_bonds}
\end{table}

The smallest cluster, which does not contain any Be-N bonds, is made up
of four \{BeO$_2$\} complexes (supercell Si$_{50}$Be$_4$O$_8$N$_{64}$,
fig. \ref{f.sibeon_struct}). As the number of Be, O atoms grows further,
the number of Si-O bonds increases linearly and equales the double
number of \{BeO$_2$\} complexes in the cell. This fact shows that the
system has no definite order because, for example, one cluster of 8
complexes \{BeO$_2$\} contains the same set of bonds as two separate
clusters each containing 4 \{BeO$_2$\} complexes (fig.
\ref{f.sibeon_struct}). As a result, atomic configurations at a definite
supercell composition may have the form of extended, compact or multiple
clusters constituted by \{BeO$_2$\} complexes, fig.
\ref{f.sibeon_struct}.

The analysis of bonds allows us to conclude that the formation of a
whole compact cluster with the maximal size is most probable. In this
case, the ``depopulation'' of FAS occurs analogously as in precipitated
phases of concentration polytypes in the AlN-Si-O system
\cite{refr,metal}, where antibonding states are present only in
``interlayer'' bonds Al-O and are absent within the ``block of
defects''.

\begin{figure}[t]
\resizebox*{\columnwidth}{!}{\rotatebox{0}{\includegraphics{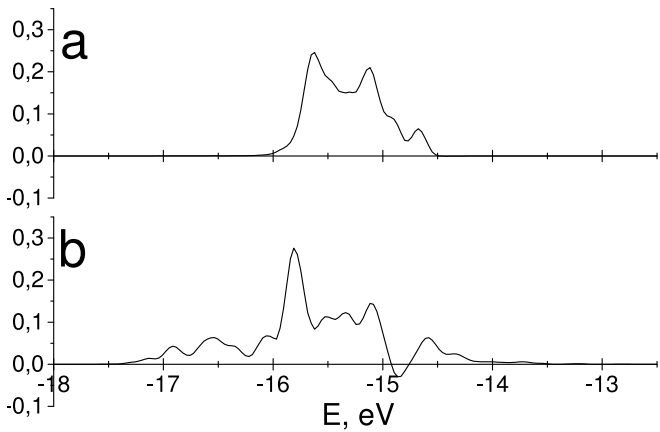}}}
\caption{COOP spectra of ``intracluster'' (a) and ``frontier'' (b) Be-O
bonds in a cluster of Be, O impurities in
Si$_{43}$Be$_{11}$O$_{22}$N$_{50}$ (fig. \ref{f.sibeon_struct}c).}
\label{f.coops_bonds}
\end{figure}

In Si$_{6-x}$Be$_x$O$_{2x}$N$_{8-2x}$, an increase in the size of the
``impurity cluster'' brings about a growth of the number of
``intracluster'' Be-O bonds reducing the number of FAS. For example,
COOP spectra of ``intracluster'' and ``frontier'' Be-O bonds of the
Si$_{46}$Be$_8$O$_{16}$N$_{56}$ supercell are depicted in fig.
\ref{f.coops_bonds}. It can be seen that intracluster bonds do not
possess any antibonding states. This indicates that the formation of a
compact cluster of the largest size is more advantageous. The growth of
a cluster in Si$_{6-x}$Be$_x$O$_{2x}$N$_{8-2x}$ is not limited meaning
the formation of a heterophase state of the system
($\beta$-Si$_3$N$_4$/BeO).

Thus, the proposed semiempirical approach based on quantum-chemical
analysis of separate interatomic bonds makes it possible to determine
the most probable effects of short-range (clusterization) and long-range
(superstructures) atomic ordering in multi-component solid solutions and
to find out the factors responsible for their stability.

Analysis of the differences between Si$_{6-x}$Be$_x$O$_{2x}$N$_{8-2x}$
and Si$_{6-x}$Al$_x$O$_x$N$_{8-x}$ solid solutions shows that the
impossibility of formation of homogeneous SS in the
$\beta$-Si$_3$N$_4$-Be-O system is due to the excess of oxygen atoms
with regard to those in $\beta$-SiAlONs. This leads to the presence of
Si-O bonds in the system, which does not allow the impurities to form
compact structures of a limited size like impurity channels. Therefore
it can be suggested that SS in the $\beta$-Si$_3$N$_4$-Be-O system may
be stabilized by altering the stoichiometry of the compounds
(composition Si$_{6-x}$Be$_x$O$_x$N$_{8-x}$) or by partial replacement
of Si atoms by elements of the IV-th group, for example, Zr
(Si$_{6-x}$Be$_{x/2}$Zr$_{x/2}$O$_x$N$_{8-x}$).

Undoubtedly, the detailed analysis of the potentially stable structures,
which were suggested in the framework of the used approach, implies
further investigations of their energy states by {\it ab-initio} methods
for the conformation of the presented results. At present, the above
studies of Si$_{6-x}$Be$_x$O$_{2x}$N$_{8-2x}$,
Si$_{6-x}$Be$_x$O$_x$N$_{8-x}$ and
Si$_{6-x}$Be$_{x/2}$Zr$_{x/2}$O$_x$N$_{8-x}$ SS are being performed by
the full-potential LMTO method.

\section{Conclusions}

In this paper we proposed a new microscopic semi-empirical method for
describing atomic ordering effects in multi-component systems. This
method is based on the analysis of interatomic bonds of separate types
and includes a procedure of minimizing the number of unfavorable bonds.

The method is used to study AOE in solid solutions of variable
compositions: Si$_{6-x}$Al$_x$O$_x$N$_{8-x}$ ($\beta$-SiAlONs) and
hypothetical  Si$_{6-x}$Be$_x$O$_{2x}$N$_{8-2x}$ in the
$\beta$-Si$_3$N$_4$-Be-O system. It is established that in $\beta$-SiAlONs
Al, O atoms form separate quasi-one-dimensional ``impurity channels''.
On the contrary, in Si$_{6-x}$Be$_x$O$_{2x}$N$_{8-2x}$ Be, O atoms trend
to form a whole compact cluster. This implies that in the
$\beta$-Si$_3$N$_4$-Be-O system homogeneous solid solutions cannot be
formed under equilibrium conditions.

Potential ways of stabilizing the SS in the $\beta$-Si$_3$N$_4$-Be-O
system are supposed to be a variation of its stoichiometry (reduction in
the oxygen concentration) or partial replacement of silicon by other
atoms from the IV-th group, for example, zirconium.

\acknowledgments

This work was supported by the Russian Foundation for Basic Research,
grant \# 01-03-96515 (Ural).


\begin{thebibliography}{}

\bibitem{gus}
A.I.Gusev, A.A.Rempel, A.A.Magerl, Disorder and Order in Strongly
Non-stoichiometric Compounds. Transition metal carbides, nitrides and
oxides, Springer, Berlin-Heidelberg, 2001.

\bibitem{order}
D.J.Lee, J.K.Lee, Acta Mater. {\bf 48}, 3847, (2000).

\bibitem{okatov2001}
S.V.Okatov, A.L.Ivanovskii, Int. J. Inorg. Mater. {\bf 3}, 923, (2001).

\bibitem{huck}
S.D.Wijevekara, R.Hoffman, Organometallics. {\bf 3}, 949, (1984).

\bibitem{ryz}
M.V.Ryzhkov, A.L.Ivanonskii, J. Struct. Chem. {\bf 43}, 18, (2002).

\bibitem{flmto}
S.V.Okatov, A.L.Ivanovskii, Phys. stat. sol. (b). {\bf 231}, R11, (2002).

\bibitem{metal}
S.V.Okatov, G.P. Shveikin, A.L.Ivanovskii, Metallofiz. Noveish. Tekhnol.
{\bf 22}, 3, (2000).

\bibitem{refr}
S.V.Okatov, A.L.Ivanovskii, G.P. Shveikin, Refract. Ind. Ceram. {\bf 41}, 270, (2000).

\end{thebibliography}
\end{document}